\documentclass[pra, twocolumn, showpacs, preprintnumbers, amsmath, amssymb, superscriptaddress, aps]{revtex4} 
\usepackage{graphicx}
\usepackage{dcolumn}
\usepackage{color}
\usepackage{bm}
\def\he4{$^4$He}
\def\hee3{$^3$He}
\def\Am3{\AA$^{-3}$}
\def\beq{\begin{equation}}
\def\eeq{\end{equation}}
\def\beqa{\begin{eqnarray}}
\def\eeqa{\end{eqnarray}}
\def\ri{\mbox{i}}
\def\rd{\mbox{d}}
\def\p{\partial}
\def\re{\mbox{e}}
\begin{document}

\author{A.B. Kuklov}
\affiliation{Department of Engineering Science and Physics,
CSI, CUNY, Staten Island, NY 10314, USA}

\author{A. Tsvelik}
\affiliation{Department of Condensed matter Physics and Materials Science, Brookhaven National Laboratory, Upton, NY 11973, USA}
\title{Quantum phase transitions to superfluid state of chains in a polarized gas of dipolar molecules}


\begin{abstract}
We analyze the nature of quantum phase transition to a superfluid state of flexible chains in a gas of polar bosonic molecules confined in
a stack of $N$ identical  1d ("cigar" type) optical lattice layers and polarized perpendicularly to the layers. Monte Carlo simulations within
the $J$-current model show that, in the absence of the inter-tube tunneling, the transition to the $N$-layered superfluid  is in the Berezinskii-Kosterlitz-Thouless universality class in the one-particle density matrix channel. The inter-layer tunneling changes it to the $q=N$ 2d Potts universality. The low energy field descriptions of the transition are discussed in terms of conformal field theories. 
\end{abstract}

\maketitle

\section{Introduction}

Quantum properties of extended objects -- high energy strings (see in Ref.\cite{strings_HE}), stripes in high-T$_c$ superconductors \cite{stripes}, 
vortices in superfluids and superconductors, dislocations in quantum crystals, {\it etc.} --  are of great interest to many areas of physics.
Recent breakthroughs in creating and trapping 
high density samples of (polar) molecules 
~\cite{Jin-Nagerl} open up the possibility of realizing \textit{quantum} chains of polar molecules.

Self-assembly of classical chains has been investigated numerically in Ref. \cite{Lozovik}. So far, quantum chains have been studied in various approximations which  neglect tunneling of particles along the chains. In Ref.\cite{Wang} it has been proposed that stiff dipolar non-interacting quantum chains should form Bose-Einstein condensate. Inter-layer pairing in bilayred dipolar systems  has been studied in Refs.\cite{bilayers}. Chains of fermionic molecules in multi-layered systems have been discussed in Refs.\cite{layers}. 

Chains of indistinguishable particles as quantum objects represent a strongly interacting system which, in general, is not amenable to the standard mean field or perturbation expansion methods. Even the second quantization for chains is generically very problematic to implement. The complexity comes from the interplay between internal degrees of freedom, that is, deformations of each chain and their centers of mass. Furthermore, several chains can exchange their segments with each other. Thus, on one side, identity of each chain is ill-defined, and, on the other, each chain looks slightly different from others due to the internal excitations \cite{note1}. Under these circumstances it is not clear what type of order quantum chains can form.   

A numerical study of quantum chains which takes into account the partial-chain exchanges as well as the intra-chain dynamics has been performed in Ref.\cite{JLTP} in the approximation of zero inter-layer tunneling. It has been shown that polar molecules in the $N$-layered geometry can form flexible (quantum rough) chains, and these chains can undergo a quantum phase transition to a superfluid phase characterized by off-diagonal long-range order (ODLRO) in the $N$-body
density matrix, while all $M$-body density matrices with $M<N$ show insulating behavior (typical for Mott insulator) regardless of the filling factor provided it is the same in each layer. If the inter-layer (dipolar) interactions are weak, a stack of $N$ layers features a $N$-component superfluid (N-SF). Once the interaction becomes stronger, the non-dissipative drag between layers will eventually convert the N-SF phase to the flexible chain superfluid (CSF) characterized by ODLRO only in the $N$-body density matrix \cite{JLTP}. The corresponding transition is continuous in the 1d-geometry and discontinuous in the 2d-geometry for $N>2$ \cite{JLTP}. The exact nature of the transition in 1d was not, however, precisely demonstrated.

\begin{figure}
\centerline{\includegraphics[angle = 0,
width=0.8\columnwidth]{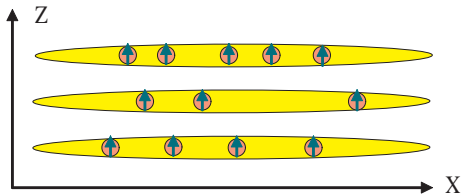}} 
\caption{(Color online)
 Schematics of 1d parallel to each other (three) layers each containing on average equal number (four) of dipole particles ---circles with arrows depicting polarization perpendicular to the layers (along $Z$-axis).
Each layer represents 1d optical lattice along $X$-axis. Tunneling along $Z$-direction is allowed, so that, while on average each layer contains the same number of particles, the relative numbers can fluctuate. The attractive interaction is mostly along $Z$-axis.} \label{pict1}
\end{figure}

In the present work we analyze how a finite inter-layer tunneling affects the transition. Since in the 2d-layers case ($N>2$) the transition is of Ist order and weak tunneling cannot change this, we concentrate on the 1d-geometry (see Fig.\ref{pict1}), that is, when layers should rather be called as tubes. 

Our main findings are as follows: The quantum phase transition into the $N$-chain superfluid is in the universality of $q=N$ 2d Potts model. That is, for $N=2,3,4$ the transition is continuous and for $N>4$ it is of first order. For $N=2$ the explicit second quantized description is presented in terms of two-component Majorana fermions and it has been conjectured that the cases $N=3,4$ can be described by, respectively, O(6) and O(8) symmetric conformal field theories. The numerically found $\nu$-exponent (of the correlation length) values are consistent with the analytical results for $N=2,3,4$. It is confirmed that the transition becomes of Ist order for $N>4$. It is also explicitly shown that the transition to CSF at zero tunneling is in the Berezinskii-Kosterlitz-Thouless (BKT) universality class.   

In the next section, we will give an argument that the transition should be in the 2d Potts universality. Then, in Sec.~\ref{Hamilton} we will describe the microscopic Hamiltonian responsible for the chain formation and will introduce the coarse-grained dual version of the Hamiltonian in the discretized time approximation. The quantities used to characterize the transition and which are evaluated numerically will be introduced. The main numerical results for the correlation length criticality will be presented as well. Finally, the effective field model will be discussed in Sec.~\ref{Majorana}. 

\section{Order parameters and universality of the transition to the CSF phase}
In the case of finite inter-layer tunneling (along $Z$ in Fig.~\ref{pict1}), the one-particle superfluid (SF) is characterized by a single phase $\varphi (x,z)$ which
defines the one-particle order parameter $\langle \psi(x,z) \rangle \sim {\rm e}^{i\varphi}$, where $\psi(x,z)$ stands for the bosonic operator and in 1d its mean is understood in terms of the algebraic order. In the CSF phase such order parameter becomes zero and the condensate remains only for the product 
\beq
\Psi (x)= \prod_{z=1,2,...,N}\psi(x,z).
\label{Psi}
\eeq
 So, the transition occurs between two superfluids: one characterized by $\langle \psi \rangle \neq 0, \, \langle \Psi \rangle \neq 0$ (SF) and the other by $\langle \psi \rangle=0, \,\, \langle \Psi \rangle \neq 0$ (CSF) (where the ODLRO in 1d is also understood in the algebraic sense). In the next section we will give more accurate definition in terms of the $M$-body density matrices.

It is important that  $\Psi$, Eq.(\ref{Psi}), is invariant with respect to the transformation 
$ \psi(x,z) \to \exp(2\pi i m(z,x)/N) \psi(x,z)$
where $m(z,x)$ is defined modulo $N$ and obeys the constraint $\sum_z m(z,x) =p N, p=0, 1, 2,... $. 
Thus, $m(z,x) $ can be broken as $m=m'+\tilde{m}$ into the discrete global part $\tilde{m}=p,\,\, p=1,2,...,N-1$ and the local gauge-type $m'(z,x)$ obeying $\sum_z m'=0$. 
Setting aside the discussion of a possible role of the local-gauge symmetry, we note that 
this discrete global transformation corresponds to the Potts model symmetry. Thus, it is natural to anticipate the $q=N$ Potts universality for the quantum SF-CSF transition.
Accordingly, for 1d tubes it should be continuous (2d Potts) for $N=2,3,4$ and discontinuous for $N>4$. The case $N=2$ corresponds effectively to pairing transition of one-component bosons. Such transition has been proposed to be in the Ising universality (which coincides with the $q=2$ Potts class) in Ref.\cite{Sachdev}. 

\section{Hamiltonian and its J-current version}\label{Hamilton}
The microscopic Hamiltonian $H$ describing SF and CSF is formulated in terms of the  creation-annihilation operators $a^\dagger_{xz},\, a_{xz}$ of a boson at site $x$ belonging to $z$th layer:
\beq
H= -  \sum_{ \langle xx'\rangle, z,z'} t_{z,z'}a^\dagger_{xz}a_{x'z'} + \frac{1}{2}\sum_{xz; x'z'}V_{xz;x'z'}n_{xz}n_{x'z'},  
\label{Habi}
\eeq
where $t_{z,z'}$ stands for a matrix of tunneling amplitudes: for $z=z'$ it is the intra-layer  tunneling amplitude $t_{||}$ between nearest-neighbors, $x'=x\pm 1$, and
for $z' \neq z$ it describes the inter-layer tunneling $t_\perp$ between sites $x=x'$ in the neighboring layers, $z'=z \pm 1$ [no tunneling between different sites located in different tubes is considered]; 
$n_{xz}=a^\dagger_{xz} a_{xz}$ denotes onsite density operator obeying the hard-core constraint; $V_{xz;x'z'}$ 
describes the matrix element for dipole-dipole interaction between sites $(xz)$ and $ (x'z')$. It is characterized
by strength $V_d=d_z^2/b^3_z$, where $d_z$ stands for the induced dipole moment and $b_z$ denotes distance between two nearest layers. This interaction is mainly attractive along $Z$ and repulsive along $X$.

\subsection{M-body density matrix}
The order can be characterized, in general, by the $M$-body density matrix
\beq
D_M =\langle \prod_{m=1,...,M}a^\dagger_{x_m,z_m} \prod_{m'=1,...,M}a_{x_{m'},z_{m'}}\rangle , 
\label{D_M}
\eeq 
where $\langle ... \rangle$ stands for the quantum-thermal averaging. 

In the 1d SF $D_1(x,z; x',z')\sim 1/|x-x'|^b, \,\, b<1,$ exhibits algebraic order at large
$|x-x'|$. In the CSF, $D_1(x,z; x',z')\sim \exp(-|x-x'|/\xi_0),\, \xi_0 \sim 1$, that is, it becomes short ranged at $T=0$ regardless of the filling factor.
Despite that, the $N$-body matrix exhibits the algebraic order. On one hand, this order is characterized by the exponential decay
$D_N(x_1,...,x_m;x'_1,...,x'_m) \sim \exp(-|x_{m_1}-x_{m_2}|/\xi_0)$ with respect to any pair of coordinates from either the set  $x_1,...,x_m$ or   $x'_1,...,x'_m$.
On the other hand, there is the algebraic order $D_N \sim 1/|R_{cm}-R'_{cm}|^c,\, c<1,$ with respect to the "center of mass" coordinates
 $R_{cm} = [x_1 + ... + x_N]/N$ and $R'_{cm} = [x'_1 + ... + x'_N]/N$ defined, respectively, for the first and the second sets of the coordinates, provided $|R_{cm} -x_m| \leq \xi_0$ and $|R'_{cm} -x'_m| \leq \xi_0$ for all $m$.

It is important to note that, while all $D_M,\, M=1,..., N-1$ exhibit algebraic ODLRO in the SF phase and exponential decay in the CSF, $D_N$ is long-ranged in both phases --- SF and CSF --- with respect to $|R_{cm}-R'_{cm}| \to \infty$ (in the algebraic sense). Thus, the transition SF to CSF can be detected by critical behavior of any $D_M, \, M<N$. We also note that the same criticality controls the long-distance behavior of $D_N$  with respect to $|R_{cm} -x_m|$ (or  $|R'_{cm} -x'_m|$). That is, in SF phase $D_N$ is trivially long-ranged with respect to $|R_{cm} -x_m|$ because $D_N$ can simply be factorized into a product of $D_1$. In contrast, in the CSF-phase, while exhibiting ODLRO with respect to $R_{cm}-R'_{cm}$, $D_N$ is short-ranged with respect to $|R_{cm} -x_m|$ (or  $|R'_{cm} -x'_m|$). Thus, the criticality can also be detected by measuring the behavior of the relative distances $x_m$ (or $x'_m$). Specifically, we have considered 
the square of so called {\it gyration radius} \cite{JLTP} as the mean of 
\beq
R^2_g=\frac{1}{N^2}\sum_{m, n=1,2,...,N}\left[ x_m - x_n\right]^2
\label{R^2}
\eeq
 with respect to the first set of the coordinates of $D_N$, provided the coordinates from the second set are kept within some distance $\sim \xi_0 $ from $R'_{cm}$ \cite{note2}.
In the SF of a length $L$, $R^2_g=R^2_{0} \sim {\cal O}(L^2) \approx \frac{1-b}{4(3-b)}L^2$, and in the CSF $R^2_{SCF} \sim {\cal O}(1) \approx \xi_0$. In what follows we will be calculating the mean of the ratio $G_N=R^2_g/R^2_0$, so that it is changing from $G_N\approx 1$ in the SF state to $G_N\sim 1/L^2 \approx 0$ in the CSF phase.
It is worth mentioning that $R_g$ can be viewed as a typical width of a chain. For strongly bound case this width is $\sim \xi_0 \approx 1$, and in the SF phase it is $\sim L$, and, thus, it exhibits critical behavior typical for correlation length.  
\begin{figure}
\centerline{\includegraphics[angle = 0,
width=1.1\columnwidth]{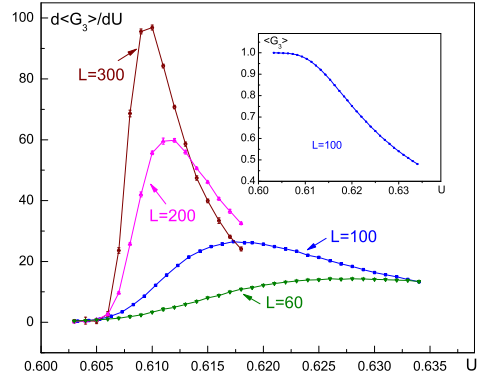}} 
\vspace{-.70cm}
\caption{(Color online)
$d\langle G_3\rangle /dU$ versus the interaction strength $U$ for $L=60,100,200,300$ with $\beta=L$. Inset: $\langle G_3\rangle$ versus $U$ for $L=100$. The SF phase corresponds to $\langle G_3\rangle\approx 1$ and the CSF to $\langle G_3\rangle\approx 0$.
The transition point SF-CSF for a given size can be identified by the maximum of $dG_3/dU$ reaching the thermodynamics limit at $U_c \approx 0.61$. 
}\label{L100}
\end{figure}
 \begin{figure}
\centerline{\includegraphics[angle = 0,
width=1.1\columnwidth]{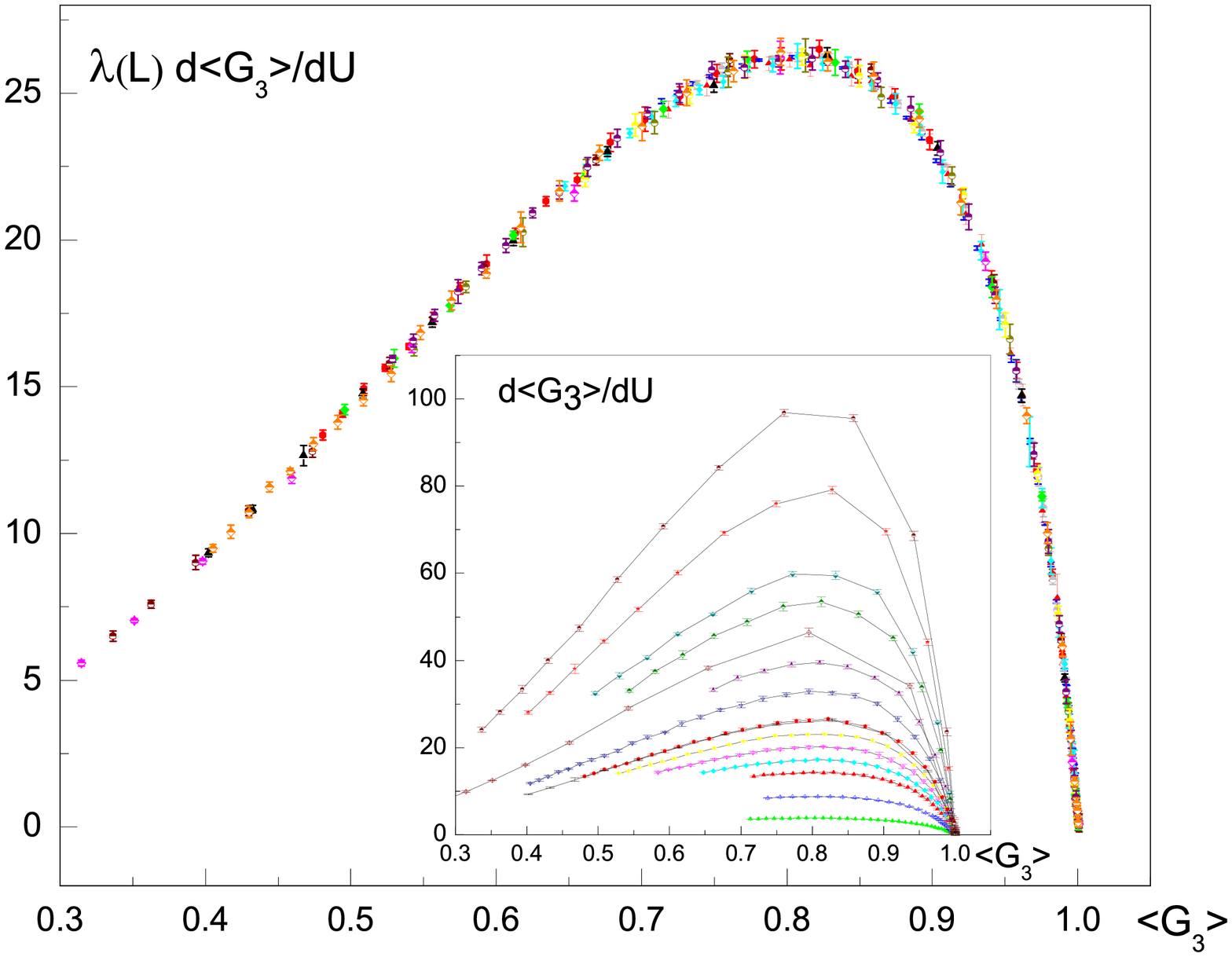}} 
\vspace{-.70cm}
\caption{(Color online)
$d\langle G_3\rangle /dU$ versus $\langle G_3\rangle$ for sizes $L=20,40, ..., 300$ rescaled by a factor $\lambda(L)$ in order to achieve collapse to the curve $L=100$ ($\lambda(100)=1$). Inset: $d\langle G_3\rangle/dU$ versus $\langle G_3\rangle$ for the same sizes.
}\label{G3}
\end{figure}
\begin{figure}
\centerline{\includegraphics[angle = 0,
width=1.1\columnwidth]{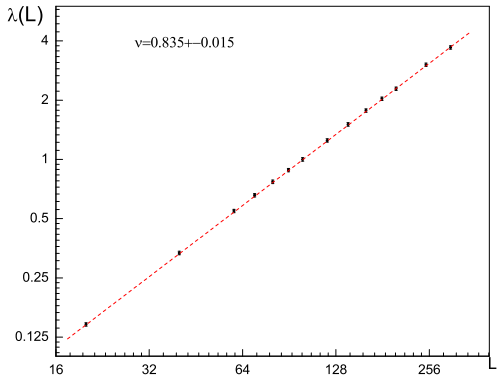}} 
\vspace{-.70cm}
\caption{(Color online)
The rescaling factor $\lambda^{-1} (L)$ versus $L$ for $N=3$ from Fig.\ref{G3}. The slope gives the correlation length exponent $\nu =0.835 \pm 0.015$. 
}\label{nu3}
\end{figure}
\begin{figure}
\centerline{\includegraphics[angle = 0,
width=0.9\columnwidth]{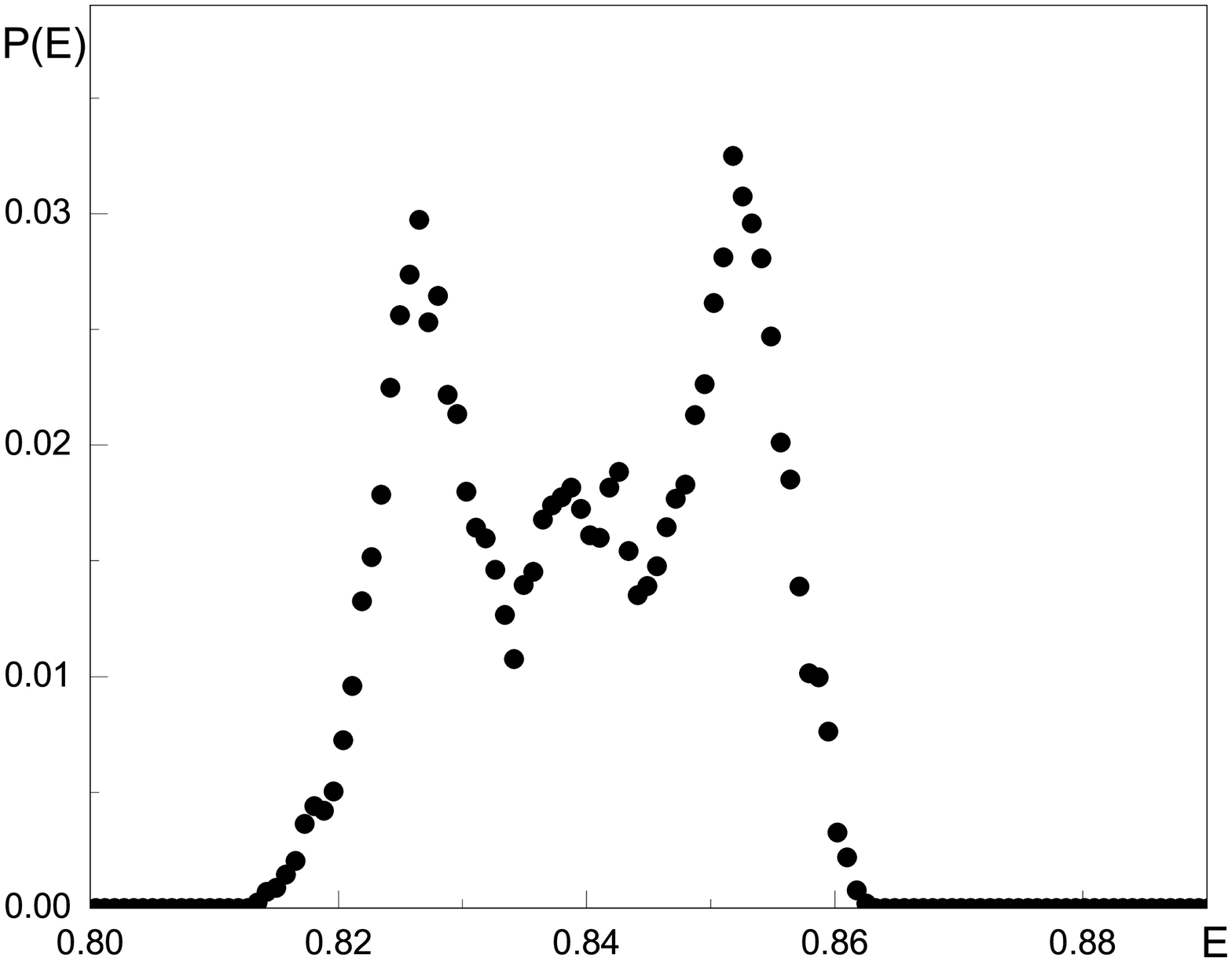}} 
\vspace{-.70cm}
\caption{(Color online)
Bimodal histogram of energy $P(E)$ for $N=5$ tubes with $L=400, \, \beta=400$. The first-order transition happens at $U_c=0.7235$. 
}\label{Iorder}
\end{figure}
\begin{figure}
\centerline{\includegraphics[angle = 0,
width=1.1\columnwidth]{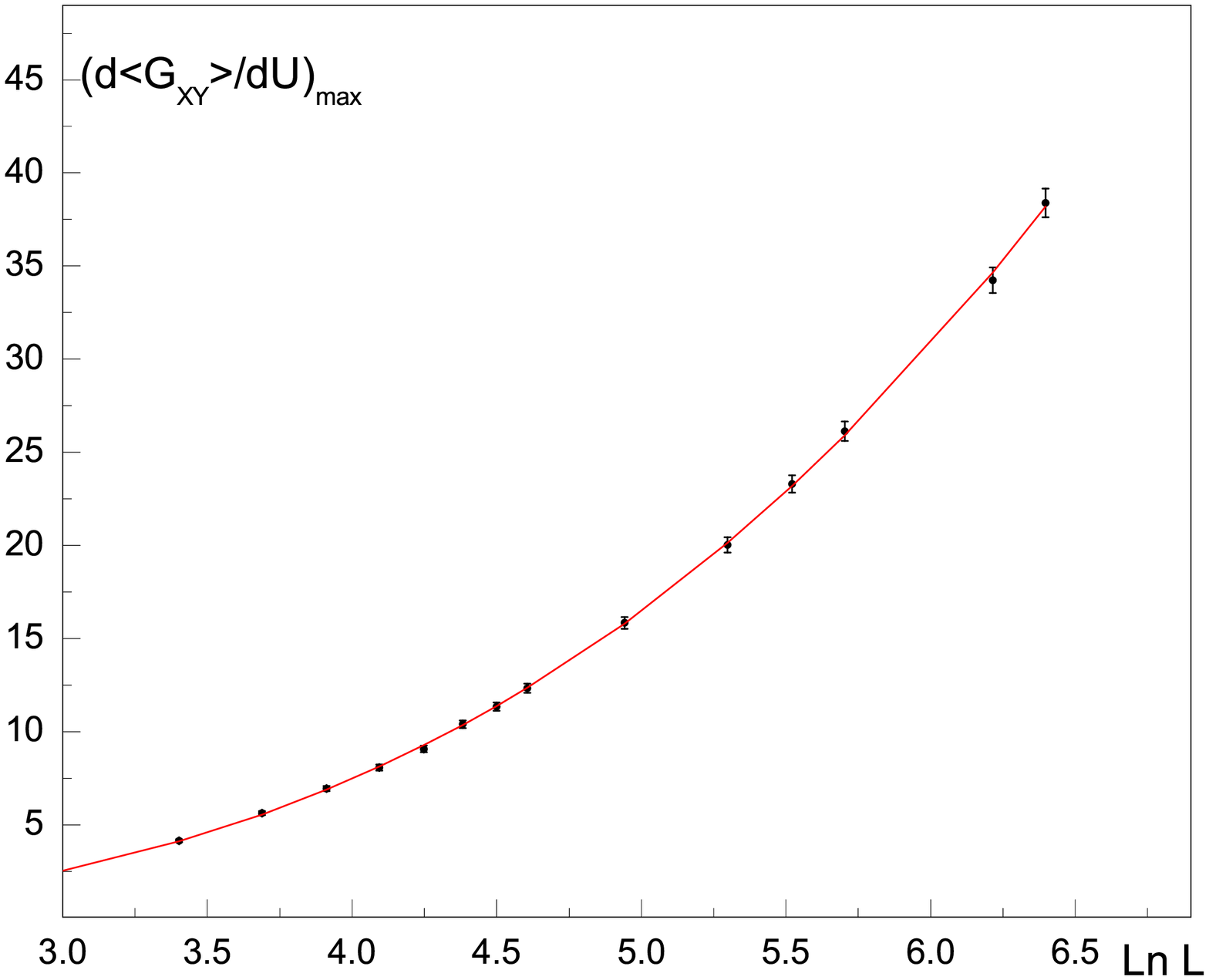}} 
\vspace{-.70cm}
\caption{(Color online)
The maximum value of $d\langle G_{XY} \rangle /dU$ versus $\ln L$ for the case of zero interlayer tunneling, $N=4$, $K(\hat{z})=\infty$. The solid red line is the fit by the
finite size scaling ansatz for the BKT-transition: $d\langle G_{XY} \rangle/dU = A \ln^3(L/L_o), \, A=0.205,\, L_o=1.98$. 
}\label{XY}
\end{figure}
\subsection{J-current formulation}\label{J-Hamilton}
Hamiltonian (\ref{Habi}) and the gyration radius (\ref{R^2}) have been used for {\it ab initio} simulations of a single chain (with exactly one polar particle per layer)  for the case $t_\perp=0$ \cite{JLTP}. It was found that the chain can undergo quantum roughening transition with the tuning parameter being the interaction strength $V_d$.
The transition is, practically, insensitive to the interaction range.  

The simulations at finite densities $n$ in each layer have been conducted in the discrete-time $J$-current-type formulation \cite{Jcurrent} of the Hamiltonian (\ref{Habi}) \cite{JLTP}. For the purpose of analyzing the universality of the transition this approach turns out to be much more efficient than the {\it ab initio} one.
Here we will be using similar model where the inter-layer tunneling is allowed. The actual dipole-dipole interaction will be replaced by onsite attraction between neighboring layers, with the intra-layer dipole-dipole repulsion ignored. The corresponding space-time action, then, becomes
\beq
H_J = \sum_{b}\left[ \frac{K(\hat{b}) (\vec{J}_{b})^{2}}{2} +\frac{U(\hat{b})(\nabla_z \vec{J}_{b})^2}{2} - \mu J^{(\hat{\tau})}_{b}\right], %
\label{HJ}
\eeq 
where $\vec{J}_{b}$ is the integer bond current obeying Kirchhoff's conservation law \cite{Jcurrent}; the summation is performed over all space-time bonds $b$ (coming out from a space-time site ($x,\tau, z$) either along $\pm \hat{x}$ or along imaginary time $\pm \hat{\tau}$ or along $\pm \hat{z}$ directions);  $\nabla_{z} \vec{J}_b \equiv \vec{J}_{b}(x,\tau,z+1) - \vec{J}_{b}(x,\tau,z)$; $\mu$ denotes chemical potential. [Here we tuned $\mu$ to have 1/2 filling of bosons per site in each tube]. Periodic boundary conditions along space $0<x<L-1, 0\leq z \leq N-1$ and along imaginary time $ 0\leq \tau \leq \beta$, with $\beta=L$, where $L=2,3, ....$, have been used. 
The coefficients $K,U$ can be related to $t_{z,z'}$ and $V_{xz;x'z'}$ from Eq.(\ref{Habi}): $K(\hat{z}) \approx 1/t_\perp$, $K(\hat{x})=K(\hat{\tau}) \approx 1/t_{||}$, 
$U\approx V_d$. The case $K(\hat{z})=\infty$ corresponds to zero inter-layer tunneling (studied in Ref.\cite{JLTP}). Here we will focus on $ K(\hat{z})= K(\hat{x})=K(\hat{\tau})$ situation as the one which naturally represents the whole universality class. 

We note that the action (\ref{HJ}) can be viewed as a coarse grained dual representation
of the Hamiltonian (\ref{Habi}). While being not precise for quantifying finite energy (non-universal) properties of the system, the J-current model \cite{Jcurrent} belongs to the same universality class as the original model (\ref{Habi}). Thus, for the purpose of this work and for sake of numerical practicality, it will be sufficient to study the model (\ref{HJ}).

We have performed Monte-Carlo simulations of the model (\ref{HJ}) within the Worm Algorithm approach \cite{WA}. 
 Green's function in imaginary time (as well as the density matrix (\ref{D_M})) is given by the statistics $D_N(\{x_m,\tau_m,z_m\};\{x'_m,\tau'_m,z'_m\})$
  of "sources" and "sinks"
of the bond currents located, respectively, at $(x_m,\tau_m,z_m),\, m=1,2,...,N$, and $(x'_m,\tau'_m,z'_m),\, m=1,2,...,N$, lattice points.
In order to insure the condition  $|R'_{cm} -x'_m| \leq \xi_0$, while  $(x_m,\tau_m,z_m),\, m=1,2,...,N$, are free to take any value,
we have convoluted $D_N(\{x_m,\tau_m,z_m\};\{x'_m,\tau'_m,z'_m\})$ with $P=\exp(-\sum_{m,n} [|x'_m-x'_n| +|\tau'_m-\tau'_n|]/\xi_0)$ as
$D_N(\{x_m,\tau_m,z_m\}; R'_{cm})=\int Dx'D\tau' Dz' D_N P \delta\left(R'_{cm} -\sum_m x'_m/N\right)$ , and, accordingly 
have evaluated the means of the normalized gyration radius $\langle G_N \rangle$ and of the center of mass distance $\langle |R_{cm}-R'_{cm}|\rangle$ where
 $\langle ... \rangle \equiv \tilde{Z}^{-1} \int Dx D\tau Dz dR'_{cm} ... D_N, \,\, \tilde{Z}= \int Dx D\tau Dz dR'_{cm} D_N$.

For sake of numerical efficiency we have symmetrized the model (\ref{HJ}) by choosing $U(\hat{b})$ independent of the type of a bond, that is, $U(\hat{b})=U$. 
The CSF phase has been identified by the condition $\langle |R_{cm}-R'_{cm}|\rangle/L =const $ and $\langle G_N\rangle \sim o(L^{-2})$ for $U>U_c$, where $U_c$ corresponds to the quantum critical point (QCP).    
In the SF phase (that is, $U<U_c$), while the first condition remained, practically, unchanged, $\langle G_N\rangle \approx 1$ with high accuracy.
The criticality of the SF-CSF transition has been analyzed through evaluating the divergent behavior of $d\langle G_N\rangle /dU$ in the vicinity of $U=U_c$.

\subsection{Finite size scaling of the gyration radius}
As discussed above, $d\langle G_N\rangle/dU$ exhibits singularity in the limit $L \to \infty$. The change from $\langle G_N\rangle \approx 1$ to $\langle G_N\rangle \approx 0$ occurs in a narrow range $\delta U= |U-U_c|$ around the critical
point $U_c \sim 1$. Such a behavior is clearly seen in Fig.~\ref{L100}:  the range of the transition $\delta U$ narrows as $L$ increases.

This range is controlled by the diverging correlation length $\xi(U) \sim |U-U_c|^{-\nu},\, \nu >0$, where $\nu$ stands for the correlation length exponent. According to the finite size scaling approach, 
$\langle G_N\rangle$ can be represented as some regular function $F(y), \, y= L/\xi(U)$ varying from $F(y=0)=1$ to $F(y=\infty)=0$ over the range $ y \sim 1$.
Thus, $d\langle G_N\rangle/dU \approx F'y /\delta U \sim L^{1/\nu}$. 
Loosely speaking, one can view this relation as $d\langle G_N\rangle/dU \approx 1/\delta U, \,\, \delta U \approx L^{-1/\nu} \to 0$. 

 We have evaluated this derivative numerically by Monte Carlo \cite{WA} and constructed 
the graphs $d\langle G_N\rangle/dU$ versus $\langle G_N\rangle$ by scanning over $U$ around the critical point $U_c$ for sizes $L=\beta = 10,20, ...300$. These graphs turn out to be self-similar so that
$d\langle G_N\rangle/dU$ for all sizes $> 10$ collapsed on a single master curve by simple rescaling of $ d\langle G_N\rangle/dU$ for size $L_1$ to another size $L_2$ as $ d\langle G_N\rangle/dU \to \lambda(L) d\langle G_N\rangle/dU$. Then, the rescaling coefficient $\lambda$, which represents the inverse width $\delta U$ as $\lambda \propto 1/\delta U$, have been plotted in the log-log$L$ axes in order to determine the critical exponent $\nu$.
The results of this procedure are presented on Figs.~\ref{G3},\ref{nu3} for the case $N=3$. [The same procedure has been used in the cases $N=2,4$ as well]. The found exponents are: $\nu=0.972 \pm 0.02$ for $N=2$, $\nu=0.835 \pm 0.015$ for $N=3$, $\nu =0.735 \pm 0.015$ for $N=4$. [The shown errors include statistical errors as well as the systematic errors due to the subleading contributions]. We note that the value of $\nu$ for $N=2$ is consistent with the $d=2$ Ising (or $q=2$ Potts) universality.  
We also note that the values of $\nu$ for $N=3$ and $N=4$ are consistent with the corresponding ones $\nu =0.837$ and $\nu = 0.756$ obtained by the Renormalization Group calculations for the $q=3,4$ 2d Potts model \cite{Dasgupta}. 

For $N>4$, the transition was found to be of first order. It has been detected by the bimodality  of energy histogram, Fig.~\ref{Iorder}. While for $N=5$ such bimodality develops on sizes $L\geq 400, \, \beta=L$, for $N=8$ it is  already  well developed  at $L=160, \, \beta=L$.  

The finite size analysis has been applied to the case of zero inter-layer tunneling as well, when
the transition is expected to be in the BKT universality. That is, $\xi \sim \exp(...|U-Uc|^{-1/2})$. The variation of the
gyration radius $\langle G_{XY}\rangle $ in this case can also be represented by some regular function $F(y)$ characterized by the range $y\sim 1$ with $y=L/\xi$ . 
Thus, $d\langle G_{XY}\rangle /dU \approx F'y \sim (\delta U)^{-3/2}\sim (\ln(L/L_o))^3$  (where $L_o$ stands for some microscopic scale) at its maximum.
The maximum value of this derivative has been plotted as a function of $L=30,...,600 $ in Fig.\ref{XY}. As can be seen the fit of the data is consistent
with the $\ln^3L $ dependence with high accuracy.

\section{The effective model in terms of Majorana fermions for $N=2$ \label{Majorana}}

Here we present the effective model describing the SF to CSF transition in the approximation where the dipole-dipole interaction
is reduced to the attraction between nearest neighbors layers. Taking into account the single occupancy constraint we can rewrite Hamiltonian (\ref{Habi})  in terms of Pauli matrix  operators:
\begin{eqnarray}
&& H - \mu N= \nonumber\\
&& \sum_j\Big\{\sum_{z=1,2}\Big[ - t_{\parallel}(\sigma^+_{j,z}\sigma^-_{j+1,z} + h.c.) + V_0\sigma^z_{j,z}\sigma^z_{j+1,z} \nonumber\\
&& - \mu\sigma^z_{j,z}\Big] + 
 t_{\perp}(\sigma^+_{j,1}\sigma^-_{j,2} +h.c.) - V_1\sigma^z_{j,1}\sigma^z_{j,2}\Big\} \label{N2}
\end{eqnarray}
where $\sigma^{\pm}$ operators stand for the bosonic creation and annihilation ones $a^{\dagger}, a$ in Eq.(\ref{Habi}) and $\hat n = \sigma^z +1$. 
While in the SF phase both fields $\sigma^-_{j,1}$ and $ \sigma^-_{j,2}$ condense (in the algebraic sense), the  CSF phase ($N=2$) corresponds to the condensation of $\Psi (j)= \sigma^-_{j,1} \sigma^-_{j,2}$, with $\sigma^-_{j,z}$ being disordered.
Another way to say this is in terms of the phases $\varphi_z$ of the fields $\sigma^-_{j,z} \sim \exp(i \varphi_z)$. The backscattering events can make $\varphi_1 - \varphi_2$ strongly fluctuating while $\varphi_1 + \varphi_2$ remains well defined. Accordingly, $\cos(\varphi_1 - \varphi_2)$ becomes irrelevant \cite{Larkin}. This constitutes the transition point from SF to $N=2$ CSF. 

In fact, similar mechanism works in mixtures of two non-convertible atomic species \cite{Mathey} (where $t_\perp=0$). It is important to note, however, that, despite such a similarity, the universality of the transition dramatically depends on $t_\perp$: while for $t_\perp=0$ it is of BKT-type \cite{Mathey}, it becomes of Ising-type at $t_\perp \neq 0$ \cite{Shulz}.

We will treat this model using bosonization technique (see in Ref.\cite{boson}) which in the context of the model (\ref{N2}) was pioneered by Schulz \cite{Shulz}. For completeness we will reproduce the calculations here. 

 The Hamiltonian (\ref{N2}) for two independent tubes in the low-energy limit (ignoring the backscattering events) is the same as for the spin S=1/2 XXZ model. 
In the continuous limit it is equivalent to the Gaussian model
 \beq
H_a = \frac{v}{2}\int \rd x \Big[K^{-1}(\p_x\Phi_a)^2 + K(\p_x\Theta_a)^2\Big], \label{Gaussian}
\eeq
 where $a=1,2$ labels the tubes and $\Theta_a$ is the field dual to $\Phi_a$: 
 $[\p_x\Theta_a (x), \Phi_a(y)] = -i\pi\delta(x-y)$. The Luttinger parameter $K$ is determined by the intra-chain interactions. If for convenience we assume that $t_{\parallel} > 0$ then the continuum limit of the operators is given by the following bosonization formulae:
\begin{widetext}
\beqa
&& \sigma^{\pm}_a(x) = \frac{1}{(2\pi)^{1/2}}\re^{\ri\sqrt{2\pi}\Phi_a} + C\Big[\re^{\ri\sqrt{2\pi}(\Theta_a + \Phi_a) +2\ri k_Fx} + \re^{\ri\sqrt{2\pi}(-\Theta_a + \Phi_a) -2\ri k_Fx}\Big] + ...\nonumber\\
  && \sigma^z_a(x) = \frac{1}{\sqrt\pi}\p_x\Theta_a + \frac{C^z}{(2\pi)^{1/2}}\sin(2k_F x + \sqrt{2\pi}\Theta_a)(-1)^n +... \label{spins}
  \eeqa 
  \end{widetext}
 where dots stand for less relevant operators and $C,C^z$ are amplitudes determined by the short range physics. 
 The Fermi momentum $k_F$ for each chain is determined by its chemical potential $\mu$ so that $k_F(\mu =0) =0$. In the following we will always assume that 
 the chemical potential is non-zero so that the spin fluctuations are incommensurate with the lattice.

Substituting (\ref{spins}) into (\ref{N2}) and defining  the fields 
\beqa
&& \Phi_{1,2} =  \sqrt\pi\Big(K^{1/2}_+\Phi_+ \pm K^{1/2}_-\Phi_-\Big), \nonumber\\
&&\Theta_{1,2} = \frac{\sqrt\pi}{2}\Big(K^{-1/2}_+\Theta_+ \pm K^{-1/2}_-\Theta_-\Big),\nonumber\\
&& [\p_x\Theta_a(x),\Phi_b(y)] = -\ri\delta_{ab}\delta(x-y),
\eeqa
 we obtain the Hamiltonian $H = H_+ + H_-$, where $H_+$ describes  the symmetric mode $(+)$: 
\beq
H_+ = \frac{v_+}{2}\int \rd x \Big[(\p_x\Phi_+)^2 + (\p_x\Theta_+)^2\Big], \label{Gaussian2}
\eeq
and $H_-$ contains only the anti-symmetric fields: 
\beqa
&& H_- = \int \rd x \Big\{ \frac{v_-}{2}\Big[(\p_x\Phi_-)^2 + (\p_x\Theta_-)^2\Big] + \label{charge2}\\
&& V_J\cos\Big(\sqrt{4\pi K_-}\Phi_- \Big)- V_c\cos\Big(\sqrt{4\pi/K_-}\Theta_-\Big)\Big\} \nonumber,
 \eeqa
 where $V_c \sim V_1, V_J \sim t_{\perp}$ and 
 \beq
 K_{\pm} = K \pm \frac{V_1}{2\pi v}.
 \eeq

At this juncture we note that the Hamiltonian (\ref{Gaussian2}) describes a mode which is not critical
at the SF to CSF transition. In other words, it is the $N=2$ CSF. In contrast, the Hamiltonian (\ref{charge2}) 
accounts for the transition so that $\Phi_-$ becomes disordered in the CSF.  

 Depending on which of the cosines in (\ref{charge2}) takes over, the ground state of this model describes either quasi long range superfluid order or pair density wave.  The latter state has a singularity in the density-density correlation function at the finite wave vector $2k_F$. 
 When both cosines are relevant (that is at $1/2 < K_- <2$) these states are separated by a QCP, the location of which is approximately determined by the relation
 $(V_J/\Lambda)^{K_-} \sim (V_c/\Lambda)^{1/K_-}$, 
 where $\Lambda$ is the ultraviolet cut-off. Fulfillment of the above condition on $K_-$ is essential for the subsequent arguments. 
 
 The vicinity of the QCP can be studied analytically when $K _-\approx 1$ (this will be our assumption throughout the rest of the paper). In that case it is convenient to refermionize (\ref{charge2}) with the result 
\begin{widetext}
  \beqa
   H_- =
  \int \rd x \Big\{ \frac{\ri v_-}{2}(-\rho_R\p_x\rho_R + \rho_L\p_x\rho_L - \eta_R\p_x\eta_R + \eta_L\p_x\eta_L) + \nonumber\\
   4\pi v_-(K_- -1)\rho_R\rho_L\eta_R\eta_L +  2\ri m_+\rho_R\rho_L + 2\ri m_-\eta_R\eta_L\Big\}, 
 \label{fin}
 \eeqa
\end{widetext}
Where $m_{\pm} = V_J \pm V_c$ and $\rho_{L,R}$ and $\eta_{L,R}$ are left- and right-moving components of Majorana (real) fermions. This model is equivalent to the continuum limit of two quantum Ising (QI) models coupled by the energy density operators \cite{Tsvelik_2011}. The Monte Carlo result  $\nu=0.972 \pm 0.02$ for the model (\ref{HJ}), $N=2$, which corresponds to the exact solution $\nu=1$ of $d=2$ Ising model, is consistent with such conclusion. 
 
In the cases $N=3,4$ we conjecture the transition to be described by, respectively, O(6) and O(8) parafermion model perturbed by the energy density
operator. The scaling dimension $\tilde{d}$ of this operator is the same as in SU$_N$(2) Wess-Zumino-Novikov-Witten model: $\tilde{d} = 4/(N+2)$ and, as consequence,  $\nu =
1/(2-d)$\cite{ZF},\cite{Q}. So, for $N=3$ we get $\tilde{d}=4/5$ and $\nu = 5/6\approx 0.833$ and for $N=4$ we get $\tilde{d}
=2/3$, that is, $\nu =0.75$. These values are consistent with the above Monte Carlo results $\nu =0.735 \pm 0.015$ for $N=4$, and  $\nu=0.835 \pm 0.015$ for $N=3$.

When this work was prepared for publication we learned about the preprint by Lecheminant and Nonne \cite{nonne} which results have a substantial overlap with ours.

\section{acknowledgements}
We are thankful to Philippe Lecheminant for useful comments.
ABK was supported by the National Science Foundation
under Grant No.PHY1005527 and by a grant of computer time from the CUNY HPCC under NSF Grants CNS-0855217 and CNS - 0958379. AMT acknowledges a support from US DOE under contract number DE-AC02-98 CH10886.


\begin{thebibliography}{99}
\bibitem{note1} Deeply in the chain phase the internal degrees of freedom can be considered as frozen in comparison with the collective dynamics of chains centers of mass.  However, close to the phase transition where all excitations are of the same order this is not the case.
\bibitem{note2} Using $D_N$ instead of $D_1$ for detecting the criticality has purely numerical advantage --- the algorithm turns out to be much more efficient.
\end{thebibliography}
\end{document}